\documentclass[doublecol]{epl2} 
\usepackage{nicefrac}
\usepackage{amsmath}

\title{Revisiting the metal-to-metal transition in $2H$-AgNiO$_2$}

\author{Jannik Gondolf, Ilya M. Eremin, Frank Lechermann}
\shortauthor{J. Gondolf et al.}

\institute{ Institut f\"ur Theoretische Physik III, Ruhr-Universit\"at Bochum,
  D-44780 Bochum, Germany                   
}
\pacs{71.27.+a}{First pacs description}
\pacs{75.40.Gb}{Second pacs description}
\pacs{74.70.Pq}{Third pacs description}

\abstract{The layered delafossite compound AgNiO$_2$ with $2H$ stacking symmetry undergoes a structural metal-to-metal transition at $T_{\rm S}\sim 365$\,K. It has been described in the past as a charge-ordering transition, where local $S=1$ spins are formed on part of the Ni sites. By means of first-principles many body calculations, we show that the transition is in fact a site-selective Mott transition on the Ni sublattice with only minor charge differentiation. Key to this finding is the uncovering of ligand-hole physics, rendering a Ni$^{2+}$ instead of a formal Ni$^{3+}$ oxidation state, in conjunction with strong local Coulomb repulsions.}

\begin{document}

\maketitle

\section{Introduction}
\begin{figure}[t]
\begin{center}
\includegraphics*[width=5cm]{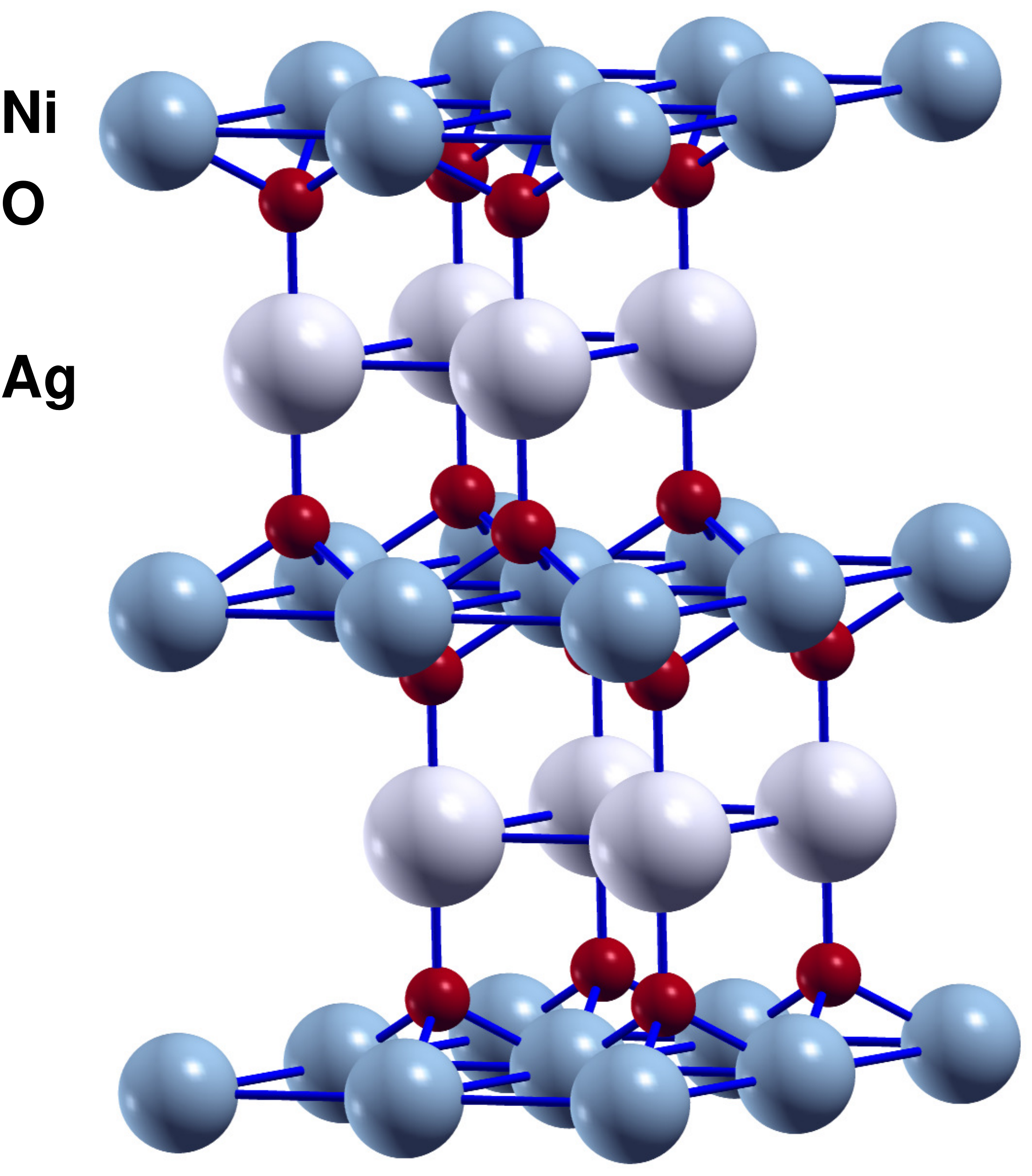}
\end{center}
\caption{Crystal structure of $2H$-AgNiO$_2$ with $P6_3/mmc$ symmetry and $c$-axis along the vertical: Ag: grey, Ni: lightblue, O: red.}\label{fig1}
\end{figure}

The physics of metallic delafossites ${\cal A}{\cal B}$O$_2$, composed of alternating layers of triangular-connected ${\cal A}$ sites and of edge-sharing ${\cal B}$O$_6$ octahedra, connected by a dumbbell O$-{\cal A}-$O bonding, has attracted considerable interest in the last decade (see e.g.~\cite{mackenzie17,daou17,lechermann-rev} for reviews). Ultrapure PdCoO$_2$ displays a surprisingly large conductivity, and PdCrO$_2$ hosts in addition layer-selective Mott physics. The silver delafossite AgNiO$_2$ is represented by two different allotropes, namely either with $R\bar{3}m$ or $2H$ stacking symmetry along the $c$ axis~\cite{soergel05}. Both structural types are conducting, but for the $2H$ compound (see Fig.\ref{fig1}a), two distinct phase transitions with temperature $T$ are reported~\cite{soergel07,waw07,waw08}. At elevated temperature, the compound is described by a high-symmetry $P6_3/mmc$ (no. 194) space group. But upon lowering $T$, a structural transition takes place at $T_{\rm S}\sim 365$\,K leading to a tripling of the in-plane unit cell from one to three formula units, resulting in a $P6_322$ (no. 182) symmetry. An antiferromagnetic transition within the Ni layers occurs furthermore at $T_{\rm N}\sim 20$\,K.

The structural transition at $T_{\rm S}$ received special attention since it connects two distinct metallic phases through a concrete in-plane lattice reconstruction. From a formal analysis of oxidation states, nickel is in a Ni$^{3+}$ state with $3d^7$ valence and hence resembles a $t_{2g}^6e_g^1$ configuration in standard symmetry-adapted orbital notation. As revealed from neutron diffraction~\cite{waw07}, the low-symmetry lattice is understood from shifts of the oxygen positions around the respective Ni sites. This gives rise to Ni1, Ni2 and Ni3 in-plane sublattices associated with the tripling of the unit cell below $T_{\rm S}$. Whilst Ni1O$_6$ octahedra expand in the low-symmetry phase, the Ni2,3O$_6$ octahedra contract.
Therefore, it was concluded~\cite{waw07}, that the given transition is most likely associated with a $\sqrt{3}\times\sqrt{3}$ charge ordering (CO)
from Ni charge disproportionation $3\,e_g^1\rightarrow e_g^2+e_g^{0.5}+e_g^{0.5}$. While the Ni1$(e_g^2)$ sites form a local $S=1$ spin from the Hund-coupled single occupation of the $d_{z^2}$ and the $d_{x^2-y^2}$ orbital, respectively, the remaining fractional filling of the Ni2,3 sites mainly ensures the metallicity in the symmetry-broken phase. This characterization of the low-symmetry phase below $T_{\rm S}$ has been supported by bond-valence sum analysis as well as measured core-level shifts in conjunction with density functional theory (DFT) data~\cite{pascut11}, yet neglecting strong correlation effects. Though the outlined picture sounds appealing, large $T$-induced charge transfers of up to one electron are usually questionable in materials~\cite{quan12}. Moreover, the formal Ni$(3d^7)$ valence often seems energetically unfavorable and replaced by a $3d^8 L$, i.e. $d^8$ with one ligand hole on oxygen, in several known nickel oxides~\cite{dem93,miz00,par12,joh14,sub15,bis16,lechermann22-2}.

In this work, a renewed theoretical investigation of the metal-to-metal transition (MMT) of $2H$-AgNiO$_2$ at $T_{\rm S}$ based on an advanced combination of DFT, self-interaction correction (SIC) and dynamical mean-field theory (DMFT) is presented. Our DFT+sicDMFT study shows that the nature of this transition is in fact of site-selective Mott kind, where the Ni1 sublattice becomes insulating and the Ni2,3 sublattices remain metallic. This finding is supported by calculations of the bare susceptibility, uncovering tendencies towards an instability of $\sqrt{3}\times\sqrt{3}$ order. The present work adds to the previous findings of site-selective Mott criticality in rare-earth nickelates~\cite{par12}, in showing that a similar mechanism may also hold for a metal-to-metal transition scenario.

\section{Theoretical Approach}
We employ the charge self-consistent~\cite{grieger12} DFT+sicDMFT framework~\cite{lechermann19}, utilizing the Ni sites as DMFT impurities, while Coulomb interactions on the oxygen sites enter by SIC on the pseudopotential level~\cite{korner10}. For the DFT part, a mixed-basis pseudopotential code~\cite{elsaesser90,lechermann02,mbpp_code} is used. Localized basis functions are deployed for Ag$(4d)$, Ni$(3d)$ and O$(2p)$. A plane-wave cutoff energy of 16 Ry is set. The SIC is applied to the O$(2s,2p)$ orbitals via weight factors $w_p$. While the $2s$ orbital is fully corrected with $w_p=1.0$, the choice~\cite{korner10,lechermann19,lechermann20-1} $w_p=0.8$ is taken for the $2p$ orbitals. Continuous-time quantum Monte Carlo in hybridization expansion~\cite{werner06} as implemented in the TRIQS code~\cite{parcollet15,seth16} solves the DMFT problem. A five-orbital general Slater-Hamiltonian, parameterized by Hubbard $U=10$\,eV and Hund exchange $J_{\rm H}=1$\,eV
\cite{lechermann20-1,lechermann2020multiorbital}, is adopted in the correlated subspace defined by Ni$(3d)$ projected-local orbitals~\cite{amadon08}. We use 36(108) Kohn-Sham bands for this latter projection in case of the high(low)-symmetry phase, starting from the bottom of the O$(2p)$-dominated states, respectively. Note that there are two-formula units associated with each primitive cell, respectively.
A double-counting correction of fully-localized-limit type~\cite{anisimov93} is used. Note that no local Coulomb interactions are applied for the Ag$(4d)$ orbitals, as in similar previous investigations of correlated delafossites~\cite{lechermann-rev}.
Analytical continuation of the Matsubara Green's functions is performed via the maximum-entropy method. The crystallographic data is taken from experiment~\cite{waw07}. For the high(low)-symmetry structure, a 13$\times$13$\times$3 (7$\times$7$\times$3) $k$-point net is utilized.

In order to get insight into possible electronic instabilities originating in the high-$T$ phase of $2H$-AgNiO$_2$ from a weak-coupling perspective, we calculate the bare susceptibility in a tight-binding framework, based on a 4$\times$4 Wannier Hamiltonian. The latter is obtained in local density approximation (LDA) by a maximally-localized Wannier construction for the low-energy DFT fourfold bands of dominant Ni-$e_g$ character. In Matsubara formalism, the bare susceptibility can thus be written in terms of LDA-based Green's functions as Lindhard term
\begin{figure}[t]
    \centering
    \includegraphics[width=0.95\linewidth]{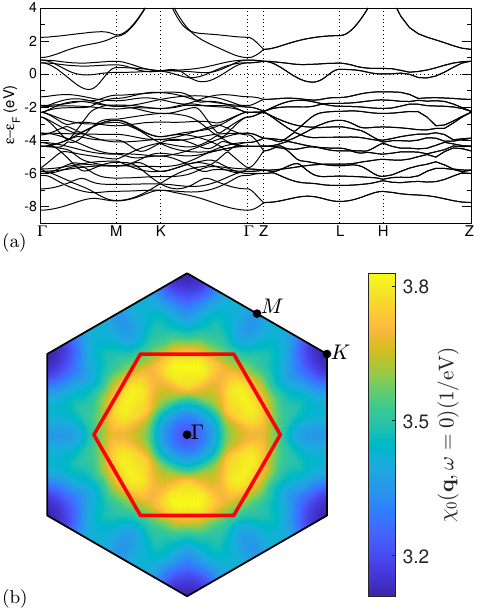}
    \caption{$\mathbf{k}$-resolved bare susceptibility of the high-symmetry phase calculated at $T=300$\,K for a $k_z=0$ cut with peaks along $\Gamma$-$K$ direction. The symmetry-broken $\sqrt{3}\times\sqrt{3}$ Brillouin zone is indicated in red.}
    \label{fig2}
\end{figure}
\begin{equation}
G_{l_1,l_2}(\mathbf{k},i\omega_n) = \sum_{\mu}  \frac{a_{l_1}^{\mu}(\mathbf{k}) a_{l_2}^{\mu*}(\mathbf{k})}{i\omega_n - \epsilon_{\mu\mathbf{k}}}\;,
\label{Eq:GF}
\end{equation}
where $a_{l}^{\mu}(\mathbf{k})$ denotes the matrix elements of the unitary transformation from orbital to band space of the non-interacting Hamiltonian. It connects the $\mu$-th band with energy $\epsilon_\mu$ with the $l$-th orbital. After performing Matsubara frequency summation and setting $i\omega_n \rightarrow \omega+i\delta$, the bare susceptibility corresponding to a bare bubble Feynman diagram follows as
\begin{equation}
\begin{split}
(\chi_0)_{l_1l_4}^{l_2l_3}(\mathbf{q},\omega+i\delta)
= \frac{1}{N}  \sum_{\mathbf{k}} \sum_{\mu\nu} 
a^{\mu}_{l_1}(\mathbf{k}+\mathbf{q}) a_{l_2}^{\mu*}(\mathbf{k}+\mathbf{q}) \\ {a}_{l_3}^{\nu}(\mathbf{k}) {a}_{l_4}^{\nu*}(\mathbf{k}) \frac{f(\epsilon_{\nu\mathbf{k}})-f(\epsilon_{\mu\mathbf{k}+\mathbf{q}})}{(\epsilon_{\mu\mathbf{k}+\mathbf{q}}-\epsilon_{\nu\mathbf{k}} )-\omega-i\delta}\;.
\label{Eq:BareSus}
\end{split}
\end{equation} 
To obtain the physically observable part of the susceptibility we contract the in- and outgoing orbital indices at the vertices of the diagram, i.e. $l_1$ with $l_4$ and $l_2$ with $l_3$. The susceptibility is calculated at zero energy $\omega=0$.
\begin{figure}[b!]
\includegraphics*[width=\linewidth]{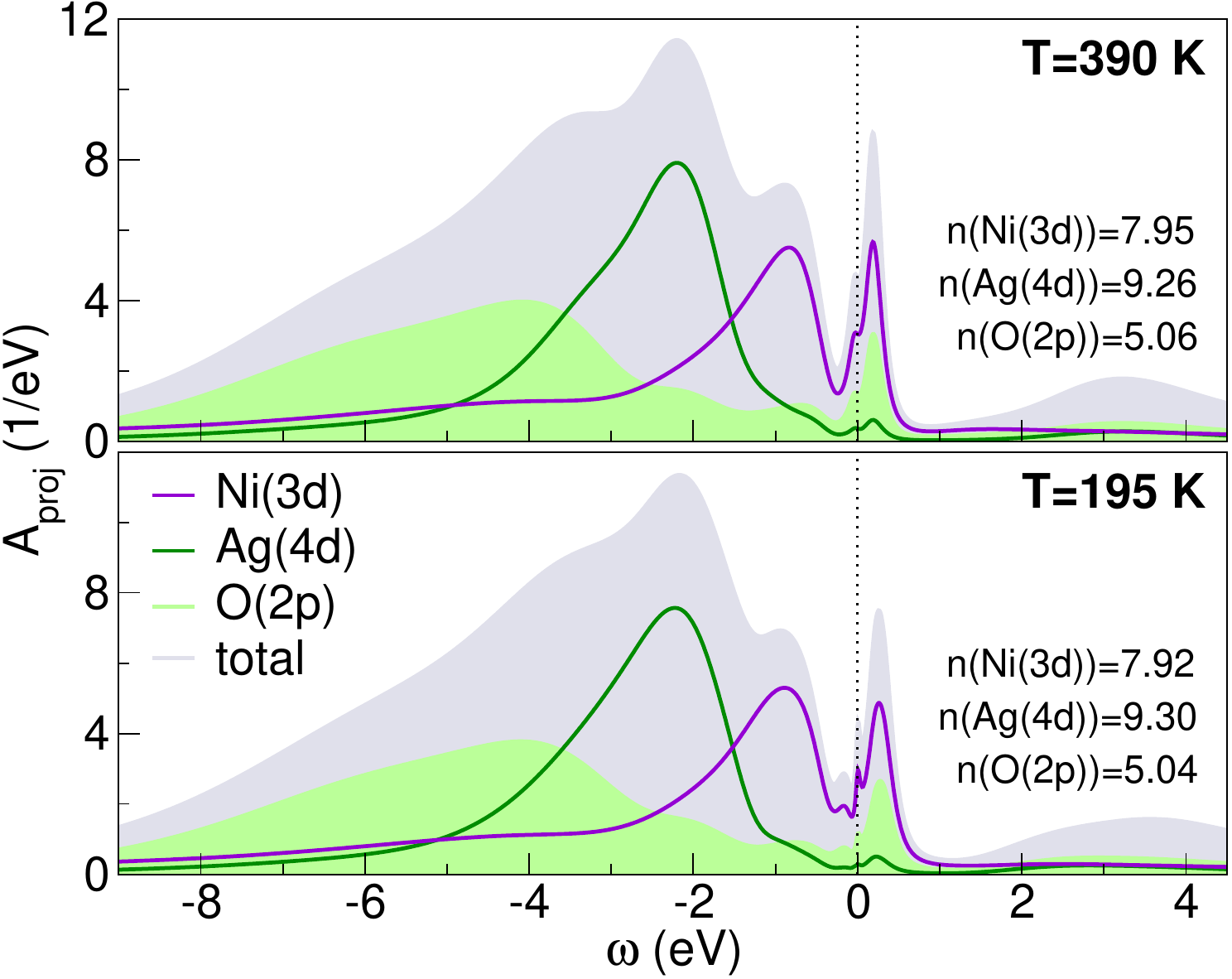}
\caption{Total and projected spectral function from DFT+sicDMFT above (top panel) and below (lower panel) the structural-transition temperature $T_{\rm S}$.}\label{fig3}
\end{figure}

\section{Results}
Let us start with the DFT electronic structure of the high-temperature structure and its susceptibility towards a possible symmetry-breaking transition. The band structure of the $P6_3/mmc$ lattice, displayed in Fig.~\ref{fig2}a, shows an isolated Ni-$e_g$ block of bands of width $W\sim 1.9$\,eV around the Fermi level $\varepsilon_{\rm F}$. Note that Ni-$d_{x^2-y^2}$ and Ni-$d_{z^2}$ are degenerate, which holds also in the low-temperature phase. The Ni-$t_{2g}$ states are merged in the wider O$(2p)$-Ag$(4d)$ block of bands in the energy window $\sim[-8,-1]$\,eV.
\begin{figure}[t]
\includegraphics*[width=\linewidth]{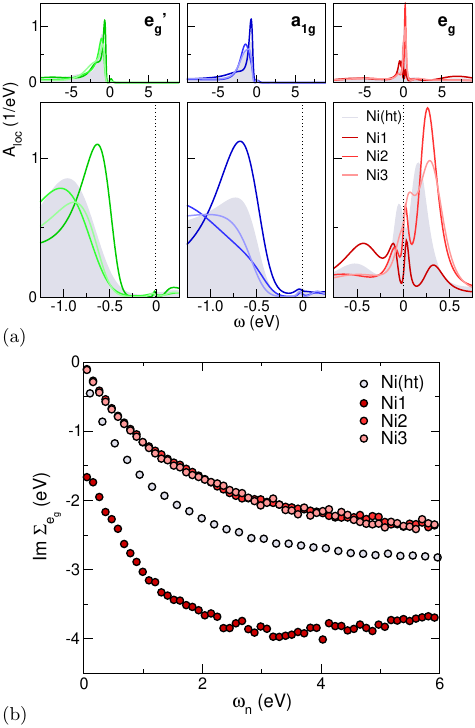}
\caption{(a) Local Ni$(3d)$ spectral function for Ni in the high-temperature (ht) phase at $T=390$\,K (grey) as well as Ni1-3 in the lower-symmetry phase at $T=195$\,K, in symmetry-adapted notation. From left to right: $t_{2g}$-based $e_g'$ (green), $a_{1g}$ (blue), and $e_g$ (red). Colors lighten from Ni1-3. Top(bottom) panel: wide(narrow) energy window around $\varepsilon_{\rm F}$.
(b) Imaginary part of the Matsubara Ni-$e_g$ self-energy for Ni1-3 as in (a), and also for Ni in the high-$T$ phase.}\label{fig4}
\end{figure}
The bare susceptibility is almost isotropic in $k_z$ direction and shows weak peaks along$\Gamma$-$K$, as seen in Fig.~\ref{fig2}b for a $k_z=0$ cut. These findings are in line with an ordering wave vector $\mathbf{k}=(1/\sqrt{3},1/\sqrt{3},0)$ \cite{waw07}, suggesting intrinsic electronic contributions to the symmetry breaking already at the level of bare (unrenormalized) quasiparticles.

To assess the differences between the metallic phases above and below $T_{\rm S}$, we performed DFT+sicDMFT calculations at $T=390$\,K and $T=195$\,K for the respective stable crystal structures. Figure~\ref{fig3} displays the resulting total as well as site- and orbital projected spectral functions $A_{\rm proj}(\omega)$. The latter are obtained from analytical continuation of the resulting Bloch Green's function and subsequent
projection via the local-orbital formalism\cite{amadon08}. Generally true for both temperatures, while the Ag$(4d)$ are very-well filled, the partially-filled Ni$(3d)$ states have a dominant contribution at the interacting $\varepsilon_{\rm F}$. Whereas these observations have been expected, surprisingly the O$(2p)$ states not only have a significant weight at low-energy, but are also far from being completely filled. The latter would be expected for a (near) O$^{2-}$ oxidation state. Instead, there is nearly exactly one (ligand) hole on each oxygen from $n($O$(2p))\sim 5$ rather than 6 for a filled $p$ shell. At the same time, Ni$(3d)$ is very close to $d^8$, and not of $d^7$ character as suggested by the formal oxidation analysis. 
Notably, the differences between the correlated electronic structure at the different temperatures appear minor in that representation. There is though some reduction of the Fermi-level total spectral weight and a sharpened Ni$(3d)$ quasiparticle (QP) peak close to zero energy, both visible in the data for the lower-temperature phase. 

Of course, the $d^8$ instead of $d^7$ filling for Ni
would speak for a half-filled Ni-$e_g$ scenario. That one is much more amenable to Mott criticality as a
$d^7$-inspired quarter filling of this orbital-symmetry sector. Let us therefore turn now to an analysis of the local Ni$(3d)$ spectral function, most interestingly for the lower-temperature phase (see Fig.~\ref{fig4}a).
The local spectral functions are directly obtained from analytical continuation of the local DMFT Green's function.
Non-surprisingly, the $t_{2g}$ orbitals of $e_g'$ and $a_{1g}$ symmetry for Ni$(3d)$ are nearly completely filled and do not play a crucial role for the low-energy physics. On the other hand, the Ni-$e_g$ orbitals, i.e. degenerate $d_{z^2}$ and $d_{x^2-y^2}$, are indeed close to half filling. In the high-temperature phase, the Ni-$e_g$ states are still well metallic, but a clear differentiation in this regard takes place in the low-temperature phase. Whereas Ni2 and Ni3 still exhibit sizeable QP-like $e_g$ spectral weight at the Fermi level, the Ni1 spectrum for $e_g$ character is (nearly) gapped. In other words, Mott criticality occurs in a site-selective way, here associated with the Ni1 site. Note that the charge differentiation among Ni1-3 remains very minor, with a formal slight enhancement in Ni1-$e_g$ of 0.05 electrons and a slight reduction of 0.05 electrons in Ni2,3-$e_g$, respectively. This hierarchy is in line with the original CO picture, yet the charge transfers here are way smaller.

The Mott criticality of the Ni1 site in the low-temperature phase becomes also obvious from inspection of the Ni-$e_g$ self-energies in Fig.~\ref{fig4}b. There, the imaginary part of $\Sigma_{e_g}(i\omega_n)$ shows strong enhancement and a very large scattering rate $\sim {\rm Im}\,\Sigma_{e_g}(i\omega_n\rightarrow 0)$ for Ni1. On the other hand, the correlation strength is apparently reduced on the Ni2,3 sites as compared to the one in the high-temperature phase. To understand the singling out of the Ni1 sublattice, remember the expanded Ni1O$_6$ octahedra in contrast to the contracted Ni2,3O$_6$ octahedra from experiment~\cite{waw07}. It leads to somewhat reduced Ni-Ni hopping from Ni1 to Ni2,3 and importantly, to higher-lying Ni1-$e_g$ levels compared to Ni2,3-$e_g$. From DFT+sic calculations, this level-energy difference amounts to 260\,meV. This enhanced Ni1-$e_g$ level energy subtly influences the site-selective charge-transfer character, eventually driving Ni1 towards Mott criticality at strong coupling. 

\section{Summary}
We re-investigated the metal-to-metal transition in $2H$-AgNiO$_2$ with realistic weak- and strong-coupling approaches. The bare susceptibility based on the DFT Ni-$e_g$ bands in the high-symmetry phase at elevated temperatures marks a tendency for symmetry breaking in the electronic subsystem towards $\sqrt{3}\times\sqrt{3}$ order already at the level of weakly-correlated itinerant electrons. Subsequent DFT+sicDMFT calculations for the high- and low-symmetry phases first reveal crucial ligand-hole physics, rendering Ni rather in the $2+$ than the formal $3+$ state throughout the whole temperature range. Secondly, the actual MMT is characterized as being of site-selective Mott kind. Whilst all of the three Ni sites in the low-symmetry unit cell display a half-filled $e_g$ manifold, only one of these sites (i.e., Ni1) becomes Mott critical. This changes the original charge-ordering scenario for $2H$-AgNiO$_2$ into a site-selective Mott scenario with only minor Ni charge differentiation. Our findings are reminiscent of what was found for the metal-insulator transition in rare-earth nickelates by Park {\sl et al.}~\cite{par12}. The difference here is however that the similar mechanism applies to a MMT in nickelates, and therefore the site-selective Mott-insulating physics may also intriguingly be realized with remaining metallicity on other Ni sublattices. Last but not least, the here proposed coexistence of Mott-critical and itinerant electrons in $2H$-AgNiO$_2$ hopefully stimulates further research. For instance, {\bf k}-resolved photoemission and/or optics may resolve novel interesting features of this system. Also the possible manipulation of the given specific nickelate delafossite physics along the lines already suggested for other metallic delafossites~\cite{lechermann-rev} would be highly interesting.

\acknowledgments
The authors are indebted to Michelle Johannes and Igor Mazin for helpful discussions.
Computations were performed at the Ruhr-University Bochum and the JUWELS Cluster of the J\"ulich Supercomputing Centre (JSC) under project miqs.

\bibliographystyle{eplbib}
\bibliography{literatur}

\end{document}